\documentstyle[epsfig,epsf]{article}
\begin{document}

\begin{center}
{\Large {\bf Magnetic Monopole Searches}}
\end{center}

\vskip .7 cm

\begin{center}
G. GIACOMELLI and L. PATRIZII \par~\par
{\it Dept of Physics, Univ. of Bologna and INFN, \\
v.le C. Berti Pichat 6/2, Bologna, I-40127, Italy\\} 

E-mail: giacomelli@bo.infn.it , patrizii@bo.infn.it

\par~\par

Lectures at the 7$^{th}$ School on Non-Accelerator Astroparticle Physics,  
\\ICTP, Trieste, Italy, 26 July - 6 August 2004. 

\vskip .7 cm
{\large \bf Abstract}\par
\end{center}

{\normalsize 
In these lecture notes we discuss  the status of the  
searches  for classical Dirac Magnetic Monopoles (MMs) at accelerators, 
for GUT superheavy MMs in the penetrating cosmic radiation 
and  for Intermediate Mass MMs. Also the searches for nuclearites and Q-balls are considered.}
\vspace{5mm}

\section{Introduction}\label{sec:intro}
The concept of magnetic monopoles (MMs) goes back to the origin of magnetism. At the beginning of the 19th century there were discussions concerning the magnetic content of matter and the possible existence of isolated magnetic charges.
In 1931 Dirac introduced the MM in order to explain the 
quantization of the electric charge~\cite{dirac}. He established the relation between 
the  elementary electric
charge $e$ and a basic magnetic charge $g$:
	$eg=n\hbar c/2= ng_{D}$,
where $n$ is an integer, $n=1,2,..$; $g_D=\hbar c/2e = 68.5 e$ is the unit Dirac charge. The existence of magnetic charges and of magnetic currents would symmetrize in form  Maxwell's equations, but the symmetry would not be perfect since $e \neq g$ (but the couplings could be energy dependent and could merge in a common value at high energies)~\cite{derujula}. There was no prediction for the MM mass; a rough estimate, obtained assuming that the classical monopole radius is equal to the classical electron radius yields  $m_M \simeq \frac{g^{2}m_e}{e^{2}} \simeq n \ 4700\  m_e \simeq n \ 2.4\  GeV/c^{2}$.
 From 1931 searches for \textit{``classical Dirac monopoles"} were carried out at every new accelerator using  simple setups, and recently also large collider detectors~$^{3-7}$.
\par
Electric charge is naturally quantized in Grand Unified Theories (GUT) of the 
basic interactions; they imply the existence of \textit{GUT monopoles} with 
calculable properties. The MMs appear in the Early Universe at 
the phase transition 
corresponding to the breaking of the unified group into subgroups, one of which is U(1)~\cite{thooft}. The  MM     
mass is related to the mass  of  the X, Y carriers of the
unified interaction, $ m_{M}\ge m_{X}/G$, 
where G is the dimensionless unified coupling constant at energies E 
$\simeq m_{X}$. 
If  
$m_{X}\simeq 10^{14}-10^{15}$ GeV and $G\simeq0.025$, $m_{M}>10^{16}-10^{17}$ GeV. 
This is an enormous
mass: MMs cannot be produced at any man--made accelerator, 
existing or conceivable. They may have been produced only in the first 
instants of 
our  Universe. \par
 Larger MM masses are expected
 if gravity is brought into the unification 
 picture, and in some  SuperSymmetric models.
\par
\textit{Multiply charged Intermediate Mass Monopoles }(IMMs) may have been produced in later
 phase transitions in the Early Universe, when a semisimple 
gauge group
yields a U(1) group~\cite{lazaride}. IMMs with m$_M$ $\sim 10^{7} \div 10^{13}$ GeV 
may be accelerated to relativistic velocities in one galactic 
magnetic field domain. Very energetic IMMs could yield the highest energy cosmic rays~\cite{bhatta}.
\par
The lowest mass MM is stable, since magnetic charge is 
conserved like electric charge. Thus the poles produced in 
the Early Universe should still exist as cosmic relics; their 
kinetic energy was affected by the  
Universe expansion and by travel through galactic and 
intergalactic magnetic fields. 
\par
GUT poles are best searched for 
underground in the penetrating cosmic radiation (CR). IMMs may be searched for at high altitude laboratories.
 \par
  In this lecture we  review the experimental situation on MM searches and briefly discuss the searches for nuclearites ~\cite{nucleariti} and Q-balls ~\cite{qballs}.

\section{Properties of magnetic monopoles}\label{sec:prop-mm}
The main properties of MMs  are obtained from the Dirac relation. \par 
\noindent - If $n$~=1 and  the basic electric charge is that of the 
electron, then  the {\it basic magnetic charge} is 
$ g_D =\hbar c/ 2e=137e/2$. The magnetic charge is larger if  $n>1$ and  if the basic electric charge is $e/3$.

\noindent - 
In analogy with the fine structure constant, $\alpha 
=e^{2}/\hbar c\simeq 
1/137$, the {\it dimensionless magnetic coupling constant} is 
$ \alpha_g=g^{2}_{D}/ \hbar c \simeq 34.25$; since it is $>1$ perturbative calculations cannot be used.
\par
\noindent - {\it Energy W acquired in a magnetic field  B}:~  
$  W = ng_{D} B\ell = n \ 20.5$ keV/G~cm.
In a coherent galactic--length   
  ($\ell\simeq 1$ kpc,  $B\simeq 3~\mu$G), the energy gained  
by a MM with $ g=g_{D}$ is
 $ W \simeq 1.8\times 10^{11}$ GeV.
 Classical poles and IMMs in the CR may 
be 
accelerated to relativistic velocities. 
   GUT poles should have low velocities, $10^{-4}<\beta<10^{-1}$. 
\par                 
\noindent- {\it MMs may be
trapped in  ferromagnetic materials}  
 by an image force, 
which  could reach  values of $\sim 10$ eV/\AA.
\par
\noindent- Electrically
charged monopoles (dyons) may arise as quantum--mechanical 
excitations or as M--p, M-nucleus composites.\par
\noindent- The interaction of a MM magnetic charge with a nuclear magnetic 
dipole 
 could lead 
to the formation of a M--nucleus bound system. 
 A monopole--proton bound state may be produced via  radiative 
capture.
Monopole--nucleus bound states may exist for nuclei with 
large gyromagnetic ratios.
 \par
\begin{figure}[ht]
	\begin{center}
		\includegraphics[width=0.8\textwidth]{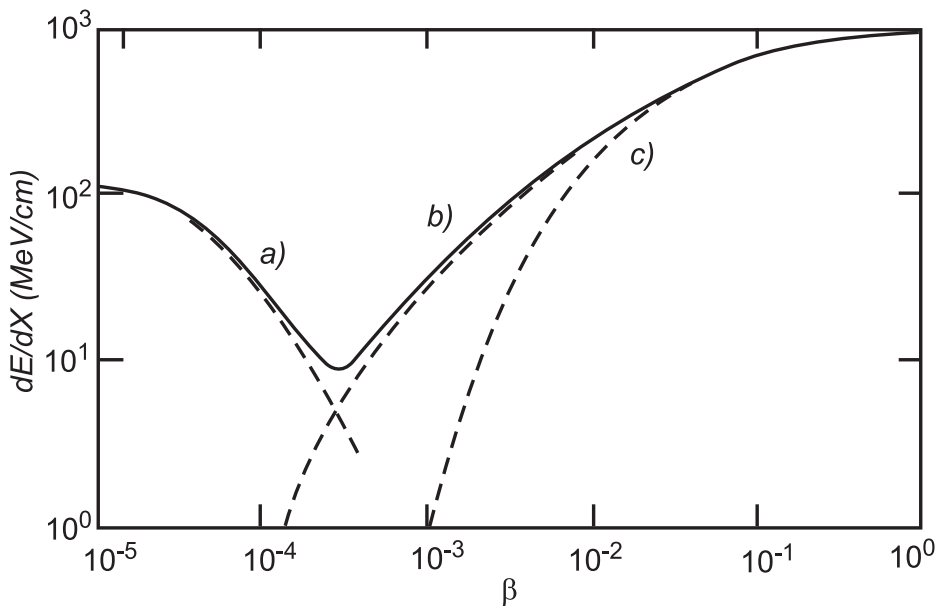}
	\end{center}
	\caption{The energy losses, in MeV/cm, of $g=g_D$ MMs in
liquid hydrogen vs ${ \beta}$. Curve a) corresponds
to elastic monopole--hydrogen atom scattering; curve b) 
to interactions with level crossings; curve c) describes
the ionization energy loss.}
	\label{fig:perdita-di-energia}
\end{figure}

 \noindent- {\it Energy losses of fast poles.} 
A fast MM with magnetic charge $g_D$ and velocity $v=\beta c$ 
behaves like an electric charge 
$(ze)_{eq}=g_D\beta$, Fig.\ \ref{fig:perdita-di-energia}.\par
\noindent - {\it Energy losses of slow poles} ($10^{-
4}<\beta<10^{-2}$) may be due to ionization or  excitation of atoms and molecules of the medium 
(``electronic'' energy
loss) or to recoiling atoms 
or 
nuclei  
(``atomic'' or ``nuclear'' energy loss). Electronic energy loss 
predominates for $\beta>10^{-3}$. 
 \par
\noindent - {\it Energy losses at very low velocities.} 
MMs with   $v<10^{-4}c$ may lose energy in elastic collisions with atoms or 
with nuclei.
 The energy is released to the 
medium in the form 
of elastic vibrations and/or infra--red radiation~\cite{derkaoui1}.\par
Fig.\ \ref{fig:perdita-di-energia} shows  the   energy loss in liquid hydrogen  
of
a $g=g_D$ MM vs  $\beta$~\cite{gg+lp}.\par

\noindent - {\it  Energy loss of MMs in celestial bodies.}
 For  $\beta$ $<10^{-4}$  the  dE/dx in 
the Earth is due to pole--atom elastic scattering, 
 eddy currents, 
and nuclear stopping power.
MMs may be stopped by celestial bodies if they have:\\ \noindent Moon: $\beta\leq 5\times {10^{-5}}$,\quad 
Earth: $\beta \leq 10^{-4}$, \quad Sun: $\beta \leq 
10^{-3}.$\par 

\section{Monopole detectors}\label{sec:mm-det}

Monopole detectors are based  on MM properties given by Dirac's 
relation. \par
\noindent - {\it  Superconducting induction devices are sensitive to MMs of any velocity~\cite{gg1}.}
A moving MM induces in a ring an electromotive force and a current change ($\Delta i$).
 For a  coil with N turns and inductance 
{\it L},  $ \Delta i=4\pi N ng_D/L=2\Delta i_o$, 
where $\Delta i_o$ is the current change corresponding to a 
change of one unit 
of the flux quantum of superconductivity. This method of
detection  is based only on the long--range 
electromagnetic interaction between the magnetic charge and the 
macroscopic 
quantum state
of a superconducting ring. \par
\noindent - {\it Scintillation counters} 
for MMs have a threshold  
$\beta \sim 10^{-4}$, above which the light signal 
is  larger than that of a minimum ionizing particle~\cite{derkaoui1,macro1}. 

\noindent - {\it Gaseous detectors } of various types have been used. 
MACRO used a gas mixture of 73\% helium and 27\% n--pentane~\cite{macro1}. This
 allows  exploitation 
of the Drell~\cite{drell} and Penning effects~\cite{gg1}: a MM leaves a  
helium atom in a metastable state (He*) with an excitation 
energy of  
$\simeq 20$ eV. The ionization potential of n--pentane is $\simeq$~10 
eV; 
the excited energy of the He* is converted  into ionization of the n--pentane 
molecule (Penning effect). \par

\noindent -
 {\it Nuclear track detectors (NTDs).}  The formation of an etchable track in a NTD is related to the Restricted
Energy Loss
(REL), the fraction of the energy loss localized
in a cylindrical region of  10 nm diameter
around the particle trajectory. It was shown that both the electronic  and the nuclear energy losses are effective in producing etchable tracks in the CR39 NTD which has a threshold at $z/\beta \simeq5$~\cite{cr39}; it is the most sensitive NTD and it  allows to search for MMs
with $g=g_D$ for $\beta$ around $10^{-4}$
and   $>10^{-3}$, the whole  $\beta$-range of
$4 \times 10^{-5}<\beta< 1$ for MMs with $g \geq 2 g_D$~\cite{derkaoui1}. The Lexan and Makrofol polycarbonates are sensitive for $z/\beta \geq 50$~\cite{barcellona}.

\section{``Classical Dirac monopoles''}
\noindent - {\it Accelerator searches.} 
If MMs are produced at high--energy accelerators, they would 
be 
 re\-la\-ti\-vi\-stic and  would ionize heavily.
 Examples of \textit{direct searches} are the experiments performed 
with scintillators or NTDs. Experiments at the Fermilab $\overline p p$ collider 
 established cross section
 limits of $\sim 2\times 10^{-34}$~cm$^2$ for MMs with $m_M<850$ GeV~\cite{bertani}.
 Searches at $e^{+}e^{-}$  colliders excluded 
masses up to 45 GeV 
 and later in the 45-102 GeV range ($\sigma<5\times 10^{-37}$~cm$^2$). Recently few high energy general purpose detectors used some subdetectors to search for Dirac MMs~\cite{opal}.\par
\begin{figure}
	\begin{center}	
		\includegraphics[width=0.8\textwidth]{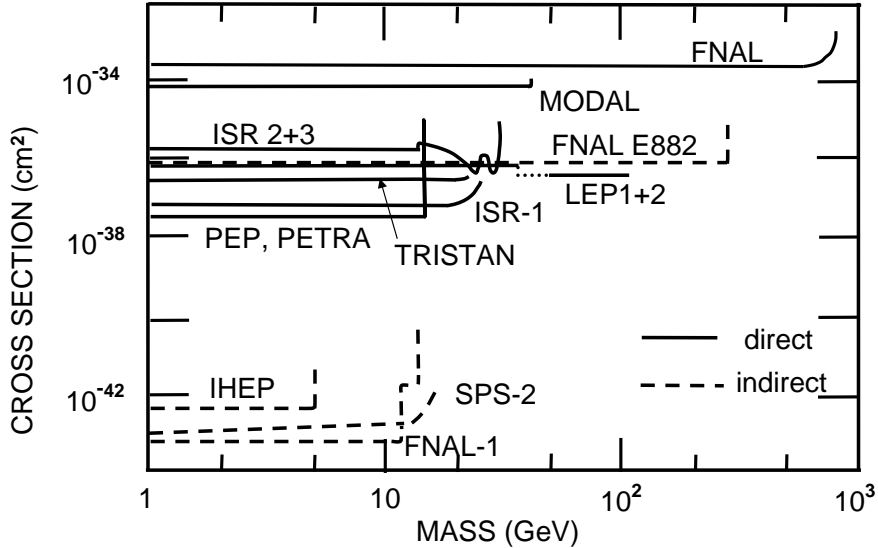}
	\end{center}
	\vspace{-5mm}
	\caption{Classical Dirac MMS cross section upper limits vs MM mass obtained 
from direct accelerator searches  (solid lines) and indirect 
searches (dashed lines).}
	\label{fig:mmclass2}
\end{figure}

Fig.\ \ref{fig:mmclass2} summarizes the cross section limits vs MM mass obtained by direct and indirect experiments (solid lines and dashed lines) at the Fermilab $\overline p p$ collider, $e^{+}e^{-}$  colliders, the ISR $p p$ collider~\cite{gg+lp}.
Most searches are sensitive to poles with magnetic charges $g =n g_{D}/q$ with $0.5<n<5$.\par
Examples of indirect searches are those performed at the CERN 
SPS and at Fermilab: the 
protons interacted  in ferromagnetic targets, 
later the targets were placed in
front of a superconducting solenoid with a field 
$B>100$ kG,  large enough to extract  and
 accelerate the MMs, to be  detected in scintillators and in  NTD  sheets~\cite{gg1}. An indirect experiment performed at the $\bar{p}p$ Tevatron collider, assumed that produced MMs could stop, be trapped and bound in the matter surrounding a collision region~\cite{kalbfleish}. Small Be and Al samples were passed through the 10 cm diameter bore of two superconducting coils, and the induced charge  measured by SQUIDs. Limits  m$_M>285$ GeV were published for $g=g_D$ poles. It is difficult to establish the validity of the hypotheses made to interpret these results.\par

\vspace{2mm}

\noindent - {\it Multi--$\gamma$ events.} 
Five peculiar photon showers found in emulsion plates 
exposed to 
high--altitude CRs, are characterized by 
an  energetic narrow cone of tens of photons, without any 
incident charged 
particle~\cite{multigamma}. The total energy of the photons is
$\sim 10^{11}$ GeV. The small radial spread of photons  suggested a c.m. $\gamma=(1-\beta^{2})^{-1/2}>10^3$. 
The energies of the photons are too small to have $\pi^o$ decays as 
their source. 
One possible explanation: a high--energy $\gamma$--ray, with energy  $>10^{12}$ eV, produced a pole--antipole pair, which suffered bremsstrahlung and annihilation producing the final multi--$\gamma$ events. 
  Searches for multi-$\gamma$ events were performed in $pp$ collisions at the ISR 
at $\sqrt{s}=53$ GeV, at the $\bar{p}p$ 1.8 TeV collider and in $e^{+}e^{-}$ collisions at LEP (Fig.\ \ref{fig:mmclass2}). The D0 experiment searched for  $\gamma$ pairs with high transverse energies; virtual pointlike MMs may rescatter pairs of nearly real photons into the final state via a box monopole diagram; they set a 95\% CL limit of 870 GeV~\cite{kalbfleish}.  At LEP the L3 coll. searched for $Z\rightarrow \gamma\gamma\gamma$ events; no deviation from QED predictions was observed, setting a 95\% CL limit of 510 GeV~\cite{kalbfleish}. Many authors studied the effects from virtual monopole loops~\cite{derujula,ginzburg}.
The authors of Ref.~\cite{anti-d0} criticized the underlying theory and believe that no significant limit can be obtained from present experiments.
\par
\noindent - {\it Searches in bulk matter.} 
 Classical MMs could be produced  by CRs and could stop at the   Earth surface, where they may be trapped in ferromagnetic materials.
Bulk matter searches used hundreds of kg of material, including meteorites, schists, ferromanganese nodules, iron ore and others. A superconducting 
coil through which the material was passed, yielded a monopole/nucleon ratio in the samples $<1.2\times 10^{-29}$ at 90\% CL~\cite{gg1}. \par
Ruzicka and Zrelov  summarized  all searches for classical poles performed before 1980~\cite{ruzicka}. A more recent bibliography is given in Ref.~\cite{biblio}. Possible effects arising from low 
mass MMs have been reported in Ref.~\cite{oscuro}.\par

\section{GUT monopoles}
\begin{figure}
	\begin{center}
		\includegraphics[width=0.74\textwidth]{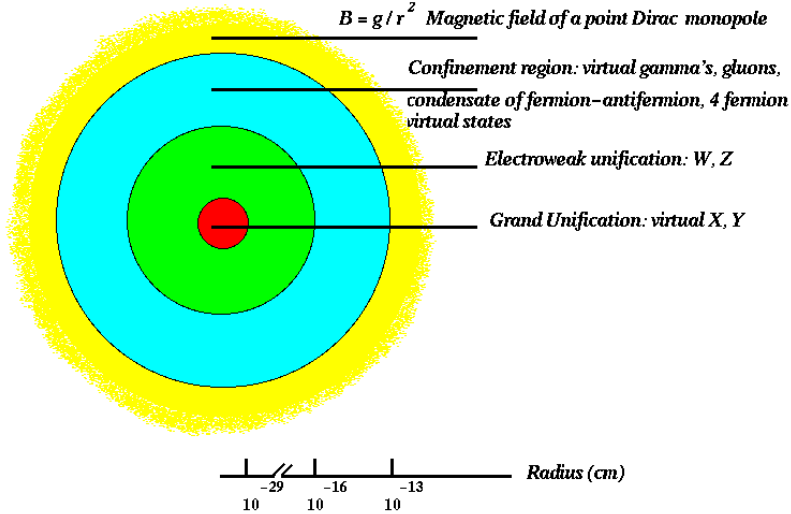}
	\end{center}
	\vspace{-3mm}
	\caption{Structure of a GUT pole. The 4 
regions correspond
to: (i) Grand Unification ($r \sim 10^{-29}$ cm; inside this core 
one finds
virtual $X$, $Y$ particles); (ii) electroweak unification
($r \sim 10^{-16}$ cm; inside one finds virtual 
$W^{\pm}$
and $Z^0$); (iii)
confinement region ($r \sim 10^{-13}$ cm; inside one finds 
virtual
$\gamma$, gluons, fermion-antifermion pairs 
and possibly 4-fermion
virtual states); (iv) for $r>$ few fm one has the field of a 
point
magnetic charge.}
	\label{fig:gut}
\end{figure}

As  already stated, GUT theories of the electroweak and strong 
interations predict the existence of superheavy MMs 
produced in the Early Universe (EU) when the
GUT gauge group breaks into separate groups, one of which is 
U(1). Assuming that the GUT group is SU(5) (which is excluded by 
proton decay experiments) one should have the following transitions:
\begin{equation}
\footnotesize
%\begin{displaymath}
    \begin{array}{ccccc}
        {} & 10^{15}\ GeV & {} & 10^{2}\ GeV & {} \\
        SU(5) & \longrightarrow & SU(3)_{C}\times \left[ SU(2)_{L}\times U(1)_{Y}\right] & \longrightarrow & SU(3)_{C}\times U(1)_{EM} \\
       {} & \small10^{-35}s & {} & \small10^{-9}s & {}
    \end{array}
%\end{displaymath}
%\normalsize
\end{equation}
MMs would be generated as topological point defects in the GUT phase transition, about
one pole for each causal domain. In the 
standard cosmology this leads to too many poles (the 
monopole problem). Inflation would defer the GUT phase 
transition after large supercooling; in its simplest version 
the number of generated MMs would be very small. However the flux depends critically on several parameters, like the pole mass, the reheating temperature, etc. If the reheating temperature is large enough one would have MMs produced in high energy collisions, like $e^{+}e^{-}\rightarrow M\bar{M}$. \\
Fig.\ \ref{fig:gut} shows the structure of a GUT MM: a very small core, an electroweak region,  a confinement region, a fermion--antifermion condensate (which may contain 4--fermion baryon--number--violating 
terms);
for $r\geq  3$ fm it behaves as a  point particle  
 generating a field $B=g/r^{2}$~\cite{picture}.

\begin{figure}
	\begin{center}
		\includegraphics[width=0.78\textwidth]{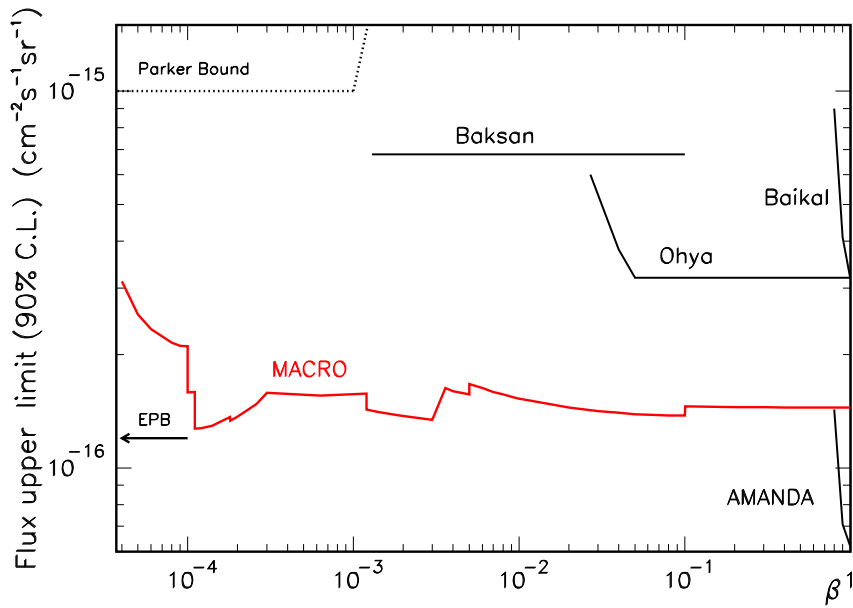}
	\end{center}
	\caption{The 90\% CL  MACRO direct upper limits vs $\beta$ for GUT  $g=g_D$ poles in the penetrating CR, and direct limits from other experiments (see text).}
	\label{fig:global2}
\end{figure}

A flux of cosmic GUT MMs may reach 
the Earth with a 
velocity
spectrum in the range $4 \times 10^{-5} 
<\beta <0.1$,
with possible peaks corresponding to the escape velocities from 
the Earth,
the Sun and the Galaxy.
Searches for such MMs in the  CR 
performed with superconducting induction
devices yielded a combined 90\%~CL limit of
$2 \times 10^{-14}~$cm$^{-2}$~s$^{-1}$~sr$^{-1}$, independent of 
$\beta$~\cite{gg+lp}.
Direct searches were performed above ground and 
underground~$^{4, 25-27}$.
MACRO  performed a search with different   types of 
detectors (liquid scintillators, limited streamer tubes and NTDs)
with an acceptance of  $\sim$ 10,000 m$^2$sr for an isotropic flux.
 No MM was detected;  the  90\% CL flux limits, shown  in
Fig.\ \ref{fig:global2} vs $\beta$  for $g=g_D$, are  at the level of $1.4\times 10^{-16}$~cm$^{-2}$~s$^{-1}$~sr$^{-1}$ for $\beta > 4 \times 10^{-5}$~\cite{mm_macro}. The figure shows also the limits from the Ohya~\cite{ohya}, Baksan, Baikal, and AMANDA experiments~\cite{baksan}.  

The interaction of the GUT monopole core with a nucleon can lead to a reaction in which the nucleon decays (monopole catalysis of nucleon decay), f. e. \( M + p \rightarrow M + e^+ + \pi^0\). The cross 
section for this process is very small, of the order of magnitude of the core size; but the catalysis process could proceed via the Rubakov-Callan mechanism with a $\sigma$ of the order  of the  strong interaction cross section~\cite{rubakov}. MACRO performed a dedicated search for nucleon decays induced by the passage of a GUT pole in the streamer tube system. The flux 
limits obtained, $3-8 \times 10^{-16}$~cm$^{-2}$~s$^{-1}$~sr$^{-1}$, depend on the MM velocity and on the catalysis cross section~\cite{catalisi}.
 Previous limits were at levels  $10^{-15}$~cm$^{-2}$~s$^{-1}$~sr$^{-1}$~\cite{catalisi}, except the Baikal limit which is $6 \times 10^{-17}$~cm$^{-2}$~s$^{-1}$~sr$^{-1}$
 for $\beta \simeq 10^{-5}$~\cite{baksan}.\par
Indirect GUT MM searches  used ancient  mica, which has a high  threshold. It is assumed that a pole passing through the Earth 
captures an 
Al nucleus and drags it through subterranean mica causing 
a trail of 
lattice defects, which survive as long as the mica is not reheated. Only small sheets were analyzed 
($13.5$ and $18$ cm$^2$), but should have been recording tracks 
for $4\div9\times 10^8$ years. The flux
limits  are 
$10^{-17} ~\mbox{cm}^{-2}~ \mbox{s}^{-1} $sr$^{-1}$ for $10^{-
4}<\beta<10^{-3}$~\cite{price}.
There are  reasons why these indirect experiments 
 might not be 
sensitive: if MMs have a positive electric 
charge  or  protons attached, then Coulomb repulsion could 
prevent capture of heavy nuclei.\par

\section{Cosmological and astrophysical bounds}
Rough upper limits for a GUT monopole flux in 
the CR 
were obtained on the basis of cosmological 
and astrophysical considerations.\par
\noindent - {\it Limit from the mass density of the universe:} 
 it is obtained requiring that the present MM mass density 
be 
smaller than the critical density $\rho_c$ of the universe. 
 For $m_M\simeq 10^{17}$ GeV one has the  
 limit:
 $F={n_Mc\over 4\pi}\beta<3\times  
10^{-12}h^2_0\beta~(\mbox{cm}^{-2}\mbox{s}^{-1} \mbox{sr}^{-1})$.
 It is valid for poles uniformely distributed in the universe. If 
poles are 
clustered in galaxies the  limit is larger~\cite{gg1}.

\noindent - {\it Limit from the galactic magnetic field (Parker limit).}
 The $\sim 3\  \mu$G magnetic field 
in our Galaxy is probably due to the non--uniform rotation of the Galaxy, 
which generates a field with a time--scale of the order of 
the rotation period of the Galaxy $(\tau\sim 10^8$ yr). An upper bound for the MM
flux is  
obtained by requiring that the kinetic energy gained per unit 
time by 
 MMs  be less than the magnetic energy generated by the dynamo 
effect: $F<10^{-
15}~\mbox{cm}^{-2}~\mbox{s}^{-1}$ sr$^{-1}$~\cite{parker}; taking into
 account  the 
almost  
chaotic nature of the field, with domains 
of 
$\ell\sim 1$ kpc, the limit becomes  mass 
dependent~\cite{parker}. 
An extended ``Parker bound", obtained by 
considering the survival of 
an early seed field~\cite{adams}, yields
$ F\leq 1.2 \times 10^{-16}(m_M/10^{17}GeV)~\mbox{cm}^{-
2}~\mbox{s}^{-1}~
\mbox{sr}^{-1}$.

\par
\noindent - {\it Limit from the intergalactic (IG) magnetic field.} If $B_{IG}\sim 3\times 10^{-8}~G$ with a 
regeneration time 
$\tau_{IG}\sim 10^9~y$, a   more stringent  bound is obtained;
 the limit is less reliable because the IG field is less 
known.
\par
\noindent - {\it Limits from peculiar A4 stars and from pulsars} may be stringent, but the 
assumptions made are not clear 
(see the pulsar PSR 1937+214)~\cite{gg1,gg+lp}.

\section{Intermediate mass magnetic monopoles}
IMMs may appear as topological point defects at a later time in the Early Universe; f.e. the SO(10) GUT group would not yield directly a U(1) group 
\begin{equation}
\footnotesize
%\begin{displaymath}
    \begin{array}{ccccc}
        {} & 10^{15}\ GeV  & {}& 10^{9}\ GeV & \\
        SO(10) & \longrightarrow & SU(4)\times SU(2)\times SU(2) & \longrightarrow & SU(3)\times SU(2)\times U(1) \\
        {} & \small10^{-35}s & {} & \small10^{-23}s & {}
    \end{array}
%\end{displaymath}
%\normalsize
\end{equation}
\noindent This would lead to MMs with masses of $\sim 10^{10}$ GeV; they would survive inflation, be stable, ``doubly charged'' ($g=2g_D$) and do not catalyze nucleon decay~\cite{lazaride}.
The structure of an IMM would be similar to that of a 
GUT 
MM, but the core would be larger (since R $\sim$ 1/$m_M$) 
and the outer 
cloud would not contain 4--fermion baryon--number--violating 
terms. \par
Relativistic IMMs, 
$10^7<m_M<10^{13}$ 
GeV, could
be present in the cosmic radiation, could  be  accelerated to large $\gamma$ in one coherent domain of the galactic field. Thus one would have to look for $\beta\ge0.1$ MMs.\par
 Detectors at the Earth surface could  detect MMs coming from above if they have $m_M>10^5-10^6$ GeV~\cite{derkaoui1}; lower mass MMs may be searched for with detectors located at high mountain altitudes,  balloons and satellites. \par
\begin{figure}
	\begin{center}
		\includegraphics[width=0.750\textwidth]{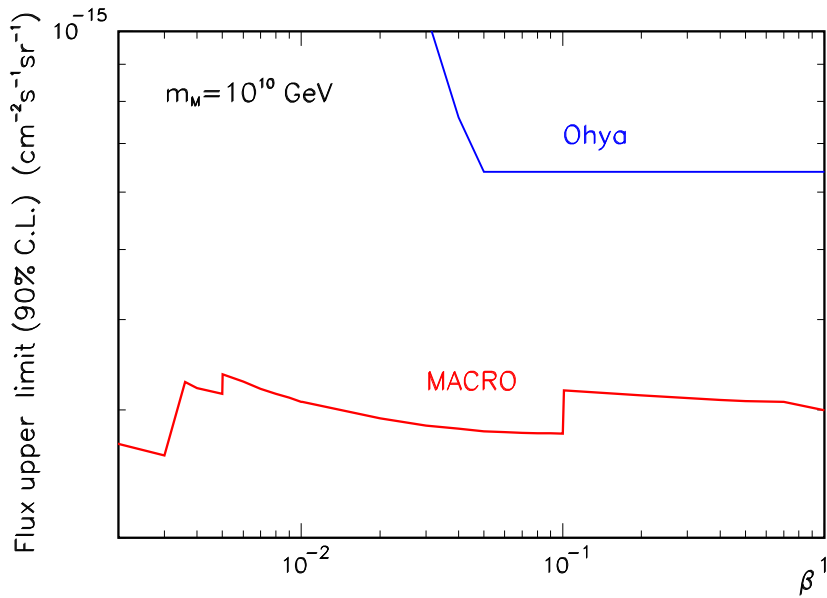}
	\end{center}
	\caption{Experimental 90\% CL upper limits for a flux of IMMs with mass $m_M=10^{10}$ GeV  plotted versus $\beta$.}
	\label{fig:imm1}
\end{figure}

 Few experimental results are available. Fig.\ \ref{fig:imm1} shows the situation on the flux upper limits for IMMs~\cite{gg+lp}.  The Cherenkov neutrino telescopes under ice and underwater are sensitive to fast ($\gamma >>1$) MMs coming from above.
 %\vspace{-10cm}

The SLIM experiment, which searches for IMMs with NTDs at the Chacaltaya
high altitude lab (5290 m a.s.l.)~\cite{slim}, is sensitive to $g=2g_D$ MMs in  the whole range $4 \times 10^{-5}<\beta <1$.

\section{Nuclearites and Q-balls}
\begin{figure}
\centerline{\epsfxsize=4.1in\epsfbox{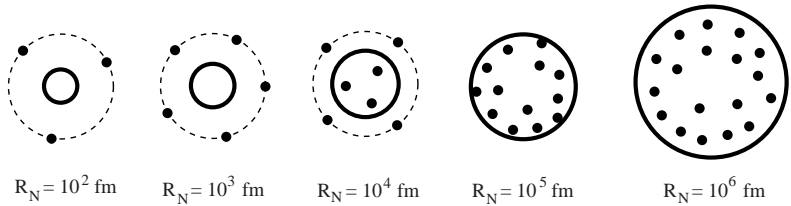}}
	\caption{Nuclearite structure. Dimensions of the quark bag (radius $R_N$) and of the core+electron system; the black points are the electrons (the border of the core~+~electron cloud for small masses is indicated by the dashed lines). For  masses smaller than $10^{9}$ GeV, the  electrons are  outside the quark bag, the core+electron system has size of $\sim 10^5$ fm; for $ 10^9 < M_N < 10^{15}$ GeV the $e^{-}$ are partially inside the core, for  $M_N>10^{15}$ GeV all electrons are inside the core. 	\label{fig:qpict}}
\end{figure}

Strange Quark Matter (SQM) should consist of aggregates of \textit{u, d} and \textit{s} quarks in almost equal proportions; the number of \textit{s} quarks should be lower than the number of \textit{u} or \textit{d} quarks and the SQM should have a positive integer charge. The overall neutrality of SMQ is ensured by an electron cloud which surrounds it, forming a sort of atom (see Fig.\ \ref{fig:qpict}). SQM should have a constant density $\rho_N = M_N /V_N\simeq 3.5 \times 10^{14}$~g~cm$^{-3}$, larger than that of atomic nuclei, and it should be stable for all baryon numbers in the range between ordinary heavy nuclei and neutron stars (A $\sim 10^{57}$).
Lumps of SQM with baryon number $A<10^6-10^7$ are usually called ``strangelets''; the word ``nuclearite'' was introduced to indicate large lumps of SQM which could  be present in the CR~\cite{nucleariti}.
SQM lumps could have been produced shortly after the Big Bang and may have survived as remnants; they could also appear in violent astrophysical processes, such as in neutron star collisions. 
 SQM could contribute to the cold dark matter.
The main energy loss mechanism for low velocity nuclearites is  elastic or quasi-elastic collisions with the ambient atoms. The energy loss  is large; therefore nuclearites should be easily detected in scintillators and CR39 NTDs~\cite{macro-nucl} .
Nuclearites should have typical galatic velocities, $\beta\sim10^{-3}$, and for masses larger than 0.1 g could traverse the earth. Most nuclearite searches were obtained as byproducts of CR MM searches; the flux limits  are similar to those  for MMs.

\begin{figure}[ht]
	\begin{center}
		\includegraphics[width=0.7\textwidth]{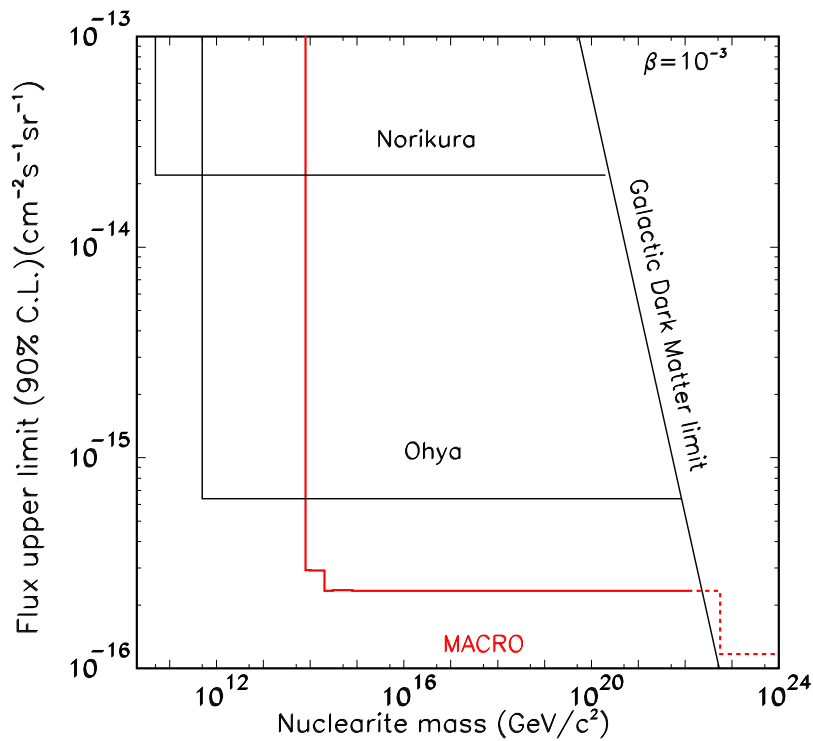}
	\end{center}
	\vspace{-5mm}
	\caption{ 90\% CL flux upper limits versus mass for nuclearites with $\beta= 10^{-3}$ at ground level. These nuclearites could have galatic or extragalatic origin. The limits are from Refs.~$^{26,35,36}$.}
\label{fig:nuclearites}
\end{figure}

The most relevant direct flux limits for nuclearites come from three large area experiments: the first two use CR39 NTDs; one experiment was performed at mountain altitude (Mt. Norikura at 2770 m a.s.l.)~\cite{nakamura}, the 2nd at the depth of $10^4$~g~cm$^{-2}$ in the Ohya mine~\cite{ohya}; the third experiment, MACRO, at an average depth of 3700 hg~cm$^{-2}$, used liquid scintillators besides NTDs~\cite{gg02}. A 4th experiment (SLIM) is deployed at high altitudes. Indirect searches with old mica samples could yield the lowest limits, but they are affected by several uncertainties. Some exotic cosmic ray events were interpreted as due to incident nuclearites, f. e. the ``Centauro'' events and the anomalous massive particles, but the interpretation is not unique~\cite{polacchi}.
Supermassive nuclearites (M $\sim$ 1 ton) passing through Earth could induce epilinear earthquakes~\cite{nucleariti,terremoti}. 
Fig.\ \ref{fig:nuclearites} shows a compilation of limits for a flux of downgoing nuclearites compared with the dark matter (DM) limit, assuming a velocity at ground level $\beta = 10^{-3}$, corresponding to nuclearites of galactic or extragalactic origin. The MACRO limit is extended above the DM bound to show the transition to an isotropic flux for $M_n>0.1$~g ($\sim 10^{23}$ GeV). Some possible positive indications are discussed in Ref.~\cite{polacchi}.

{\it Q-balls} should be aggregates of squarks $\tilde{q}$, sleptons $\tilde {l}$ and Higgs fields~\cite{qballs}. The scalar condensate inside a Q-ball core has a global baryon number Q (and may be also a lepton number). Protons, neutrons and may be electrons could be absorbed in the condensate.
There could exist neutral and charged Q-balls. Supersymmetric Electrically Neutral Solitons (SENS) are generally massive and may catalyse proton decay. SENS may obtain a positive electric charge absorbing a proton in their interactions with matter yielding SECS (Supersymmetric Electrically Charged Solitons), which have a core electric charge, have generally lower masses and the Coulomb barrier could prevent the capture of nuclei. SECS have only integer charges because they are color singlets.
A SENS which enters the earth atmosphere could absorb a nitrogen nucleus which would give it the positive charge of +7 (SECS with $z=7$). Other nuclear absorptions are prevented by Coulomb repulsion. If the Q-ball can absorb electrons at the same rate as protons, the positive charge of the absorbed nucleus may be neutralized by the charge of absorbed electrons. If, instead, the absorption of electrons is slow or impossible, the Q-ball carries a positive electric charge after the capture of the first nucleus in the atmosphere.
Q-balls may be cold DM candidates. SECS with $\beta \simeq 10 ^{-3}$ and  $M_Q < 10^{13}$ GeV could reach an underground detector from above, SENS also from below. SENS may be detected by their  continuons emission of charged pions (energy loss  $\sim$ 100 GeV g$^{-1}$cm$^2$), SECS may be detected by scintillators, NTDs and  ionization detectors.\par
Note that we did not consider here the possibility of strongly interacting, colored, MMs, nuclearites~\cite{wick} and Q-balls. 

\section{Conclusions. Outlook}
Direct and indirect accelerator searches for classical Dirac MMs
placed  
limits at the level  $m_M > 850$ GeV with cross section upper values as shown in Fig.\ \ref{fig:mmclass2}. Future improvements may
come from
experiments at the LHC~\cite{moedal}.
\\
\indent Many searches were performed for GUT poles in the penetrating cosmic radiation. The 90\% CL flux limits are at $\sim
1.4 \times 10^{-16} $~cm$^{-2}$~s$^{-1}$~sr$^{-1}$ for $\beta \ge 
4 \times 10^{-5}$.
It may be difficult to do much better since one would require refined
detectors of
considerably larger areas.
\par
Present limits on Intermediate Mass Monopoles with high $\beta$ are relatively poor. 
Experiments at high altitudes and at neutrino telescopes should
 improve  the situation. In particular stringent limits may be obtained by large neutrino telescopes for IMMs with $\beta > 0.5$ coming from above. \par
As a byproduct of GUT MM searches some experiments obtained stringent limits on nuclearites and on Q-balls. Future experiments at neutrino telescopes and at high altitudes should perform searches for nuclearites and Q-balls of smaller masses.

\section{Acknowledgements}
We acknowledge the cooperation of many colleagues, in particular S. Cecchini, M. Cozzi, M. Giorgini, G. Mandrioli, S. Manzoor, V. Popa, M. Spurio, and others. We thank ms. Giulia Grandi for typing the manuscript.


\begin{thebibliography}{99}
\bibitem{dirac} P.A.M. Dirac, Proc.~R.~Soc. London 133(1931)60; Phys. Rev. 74(1948)817.
\bibitem{derujula} A. De Rujula, Nucl. Phys. B435(1995)257.
\bibitem{gg1}G. Giacomelli, Riv. Nuovo Cimento 7(1984)N.12, 1.
\bibitem{gg+lp}G. Giacomelli et al. hep-ex/011209; hep-ex/0302011; hep-ex/0211035.
\bibitem{kalbfleish} G.R. Kalbfleisch, Phys. Rev. Lett. 85(2000)5292. K.A.Milton et al. hep-ex/0009003. B. Abbott et al., Phys. Rev. Lett. 81(1998)524. M. Acciarri et al., Phys. Lett. B345(1995)609.
\bibitem{anti-d0} L. Gamberg et al., hep-ph/9906526.
\bibitem{opal} Private communication by M. Cozzi. \par\noindent K. Kinoshita et al., Phys. Rev. D46(1992)R881.
\bibitem{thooft} G.'t Hooft, Nucl. Phys. B29(1974)276. A.M. Polyakov, JETP Lett. 20(1974)194.
N.S. Craigie et al., Theory and Detection of MMs in Gauge Theories, World Scientific, 
Singapore (1986).
\bibitem{lazaride} G. Lazarides et al., Phys. Rev. Lett. 58(1987)1707.\par\noindent
T. W. Kephart and Q. Shafi, Phys. Lett. B520(2001)313.
\bibitem{bhatta} P. Bhattacharjee and G. Sigl, Phys. Rept. 327(2000)109 and refs. therein.
\bibitem{nucleariti} E. Witten, Phys. Rev. D30(1984)272.\par\noindent A. De Rujula and S. Glashow, Nature 31(1984)272.
\bibitem{qballs} S. Coleman, Nucl. Phys. B262(1985)293. \par\noindent A. Kusenko and A. Shaposhnikov, Phys. Lett. B418(1998)46.
\bibitem{derkaoui1} J. Derkaoui et al., Astrop. Phys. 9(1998)173; Astrop. Phys. 9(1999)339.
\bibitem{macro1} S. Ahlen et al., Phys. Rev. Lett. 72(1994)608. M. Ambrosio et al., Astrop. Phys. 6(1997)113; Nucl. Instr. Meth. A486(2002)663; Astrop. Phys. 4(1995)33; Astrop. Phys. 18(2002)27.
\bibitem{drell} G.F. Drell et al., Nucl. Phys. B209(1982)45.
\bibitem{cr39} S. Cecchini et al., Nuovo Cim. A109(1996)1119. 
\bibitem{barcellona} S. Cecchini et al., 22th ICNTS, Barcelona, Spain, 2004, hep-ex/0503003; hep-ex/0502034.
\bibitem{bertani} M. Bertani et al., Europhys. Lett. 12(1990)613.
\bibitem{multigamma} M. Schein et al., Phys. Rev. 99(1955)643.
\bibitem{ginzburg}I. F. Ginzburg and A. Schiller, Phys. Rev. D60(1999)075016.
\bibitem{ruzicka} J. Ruzicka and V.P. Zrelov JINR-1-2-80-850(1980).
\bibitem{biblio} G. Giacomelli et al., hep-ex/0005041.
\bibitem{oscuro} V.A. Skvortsov et al., 29th EPS Plasma Conf., ECA 26B, D--5.013 (2002).
\bibitem{picture} D. Bakari et al., hep-ex/0004019. 
\bibitem{mm_macro} M. Ambrosio et al., MACRO Coll., Eur. Phys. J. C25(2002)511; Phys. Lett. B406(1997)249; Phys. Rev. Lett. 72(1994)608.
\bibitem{ohya}  S. Orito et al. (\lq\lq Ohya''), Phys. Rev. Lett. 66(1991)1951. 
\bibitem{baksan}E.N. Alexeyev et al. (\lq\lq Baksan''), 21$^{st}$ ICRC 
10(1990)83.	\par\noindent V.A. Balkanov et al. (\lq\lq Baikal'') Nucl. Phys. B(Proc. Suppl.) 91(2001)438. \par\noindent P.Niessen et al., 27$^{st}$ ICRC 3(2001)1496.	
\bibitem{rubakov} V.A. Rubakov, JETP Lett. B219(1981)644. \par\noindent G.G. Callan, Phys. Rev. D26(1982)2058.
\bibitem{catalisi} M. Ambrosio et al., Eur. Phys. J. C26(2002)163.
\bibitem{price} P. B. Price, Phys. Rev. D38(1988)3813. \par\noindent D. Ghosh and 
S. Chatterjea, Europhys. Lett. 12(1990)25.
\bibitem{parker} E.N. Parker, Ap. J. 160(1970)383. \par\noindent M.S. 
Turner et al., Phys. Rev. D26(1982)1296.
\bibitem{adams}F.C. Adams et al., Phys. Rev. Lett. 70(1993)2511.
\bibitem{slim} D. Bakari et al., hep-ex/0003028. S. Cecchini et al. 28$^{st}$ ICRC 3(2003)1657; Nucl. Phys. B(Proc. Suppl.)138(2005)529.
\bibitem{macro-nucl} M. Ambrosio et al., Eur. Phys. J. C13(2000)453.
\bibitem{nakamura} S. Nakamura et al., Phys. Lett. B263(1991)529.
\bibitem{gg02} G. Giacomelli, hep-ex/0210021.
\bibitem{polacchi} M. Rybczynski et al., hep-ph/0410064
\bibitem{terremoti} D. P. Anderson et al., astro-ph/0205089
\bibitem{qball-enlos} D. Bakari et al., Astrop. Phys. 15(2001)137.
\bibitem{arafune} J. Arafune et al., hep-ph/0005103.
\bibitem{wick}S.D. Wick et al., astro-ph/0001233.
\bibitem{moedal} Proposal MOEDAL at the LHC, CERN/LHCC 98-5.
\end{thebibliography}
\end{document}